\documentclass[11pt]{article}
\usepackage[left=1.5in,top=1.5in,right=1.5in,bottom=1.25in]{geometry}
\usepackage{amsmath}
\usepackage{amssymb}
\usepackage{amsbsy}
\usepackage{amsthm}
\usepackage{epsfig}
\usepackage[round]{natbib}
\usepackage{clrscode}
\usepackage{setspace}
\usepackage{multirow}

\usepackage{mathptmx}

\newcommand{\N}{\mbox{N}}

\newcommand{\C}{\mbox{C}}
\newcommand{\Be}{\mbox{Be}}
\newcommand{\dd}{\mbox{d}}

\newcommand{\Ga}{\mbox{Ga}}

\newcommand{\Ex}{\mbox{Exp}}
\newcommand{\E}{\mbox{E}}

\newcommand{\IB}{\mbox{IB}}
\newcommand{\Z}{\mbox{Z}}
\newcommand{\GZ}{\mbox{GZ}}
\newcommand{\Meix}{\mbox{Meix}}

\newcommand{\IN}{\mbox{IN}}

\newcommand{\bbeta}{\boldsymbol{\beta}}
\newcommand{\balpha}{\boldsymbol{\alpha}}
\newcommand{\btheta}{\boldsymbol{\theta}}

\newcommand{\by}{\mathbf{y}}
\newcommand{\p}{p}

\newcommand{\beps}{\boldsymbol{\epsilon}}

\newtheorem{theorem}{Theorem}

\title{Local Shrinkage Rules, L\'evy Processes, and Regularized Regression}

\author{
\textsc{Nicholas G.~Polson} \\
\small{\textit{Booth School of Business}} \\ \small{\textit{University of Chicago}} \\
\\
\textsc{James G.~Scott} \\
\small{\textit{Division of Statistics and Scientific Computing}} \\
\small{\textit{and McCombs School of Business}} \\ \small{\textit{University of Texas at Austin}} \\
}

\date{Original version: June 2010 \\
Revised: April 2011}
\begin{document}

\maketitle
\begin{abstract}
We use L\'evy processes to generate joint prior distributions, and therefore penalty functions, for a location parameter 
$\bbeta = ( \beta_1 , \ldots , \beta_p ) $ as $p$ grows large. This generalizes the class
of local-global shrinkage rules based on scale mixtures of normals, illuminates new connections among disparate methods, and leads to new results for computing posterior means and modes under a wide class of priors. We extend this framework to large-scale regularized
regression problems where $p>n$, and provide comparisons with other methodologies.

\vspace{0.1in}
\end{abstract}


\section{Introduction}

In recent years there has been considerable interest in the subject of choosing a joint prior distribution, or equivalently a penalty function, for a high-dimensional location vector $\bbeta = (\beta_1, \ldots, \beta_p)$.  Much of this work has been motivated by problems in high-dimensional regularized regression, where we observe data $\by = X \bbeta + \beps$ and wish to estimate $\bbeta$.  Good examples of recent Bayesian research in this area are the papers of \citet{park:casella:2008} and \citet{hans:2008}, who explore Bayesian versions of the traditional lasso penalty.

In the case where the number of regressors is moderate, the use of normal variance mixtures to generate exchangeable joint distributions for $\bbeta$ has been studied in detail.  Such priors arise from a hierarchical model where
\begin{eqnarray}
(\beta_j \mid \tau^2, \lambda_j^2) &\sim& \N(0, \tau^2 \lambda_j^2) \label{global-local1} \\
\lambda_j^2 &\sim& \p(\lambda_j^2) \label{global-local2}  \\
(\tau^2, \sigma^2) &\sim& \p(\tau^2, \sigma^2)  \label{global-local3}  \, ,
\end{eqnarray}
with the $\lambda_j^2$'s known as the local shrinkage parameters.  The following section reviews several recent proposals along these lines.

The analogous classical formulation is to identify a penalty function with the log prior:
\begin{eqnarray*}
g(\beta_j) &=& - \log p(\beta_j \mid \tau^2) \\
p(\beta_j \mid \tau^2) &=& \int_0^{\infty} p(\beta_j \mid \tau^2, \lambda^2) \ p(\lambda_j^2) \ d \lambda_j^2 \, .
\end{eqnarray*}
Upon observing data $\by$, $\bbeta$ is chosen to minimize
\begin{equation}
\label{pl.objective.equation}
l(\bbeta) = \Vert \by - X \bbeta \Vert^2 + \nu \sum_{i=1}^p g( \beta_j ) \, ,
\end{equation}
where $\tau^2$ has been re-expressed in terms of the regularization parameter $\nu$.  The minimizing choice of $\bbeta$ is equivalent to the joint posterior mode under the Bayesian formulation.

Our interest in this framework arises from the intersection of two challenges, one theoretical and the other practical.
\begin{enumerate}
\item The usual variance-mixture approach provides an insufficiently general understanding of how priors, penalty functions, and Bayesian variable selection are related.  For example, some penalty functions lack variance-mixture equivalents \citep[e.g.][]{fan:li:2001}, while some variance mixtures lead to penalty functions without closed-form representations \citep[e.g.][]{Carvalho:Polson:Scott:2008a}.  Meanwhile, Bayesian variable selection, involving a complex prior and a large discrete space of submodels, would seem a world apart from either approach.
\item As $p$ grows large, many potentially useful approaches become computationally intractable, or yield answers whose quality is difficult to assess using standard tools.
\end{enumerate}

In this paper, we explore an alternative to the traditional framework by constructing joint priors for $\bbeta$ using L\'evy processes.  This provides a unifying probabilistic structure for penalized regression and variable selection from both Bayesian and classical viewpoints.

Even within this new framework, a complete theory capable of completely solving both challenges listed above remains beyond our reach.  Nonetheless, we will argue that the L\`evy-process view offers several insights that are highly relevant to statistical practice.  First, our approach embeds finite-dimensional normal variance mixtures in a wider class of infinite-dimensional, non-Gaussian joint distributions.  It therefore provides an intuitive framework for asymptotic analysis on existing priors and penalties, as well as a device for generating previously unexplored options (such as the Meixner and $z$-distributions discussed in Section \ref{sec:levyprocess}).  The use of L\'evy processes in high-dimensional Bayesian modeling has been gaining in popularity \citep[e.g.][]{wolp:taqq:2005,Wolp:Clyd:Tu:2010}.  Our approach differs from this line of work, in that we wish to use the theory of L\'evy processes to provide a general framework of penalty functions, shrinkage priors with exchangeable structure, and the relationship between them.  Sections \ref{sec:levyprocess} and \ref{sec:generallevycase} will explore these relationships in depth, while Section \ref{sec:statisticalrel} will demonstrate their statistical relevance.

Second, we show that both Bayesian variable selection and the pure-shrinkage approach of something like the lasso can be subsumed into a unified theoretical framework.  Connections between these two approaches are important due to the acute computational difficulties associated with the high-dimensional variable-selection problem.  Indeed, Section \ref{sec:statisticalrel} describes an asymptotic sense in which the two models agree on certain important features.

Finally, our framework provides new insight on how the two quantities most typically of interest---the posterior mode and mean of $\bbeta$---can be computed.  We prove a theorem characterizing the posterior mean for $\bbeta$ in terms of the L\'evy measure of the subordinator used to construct the joint prior $p(\bbeta)$.  This theorem can be used to understand the issue of Bayesian robustness for a much wider class of priors than those for which existing tools are sufficient \citep[c.f.][]{pericchi:smith:1992,griffin:brown:2010}.  We also show how the L\'evy-process approach leads to a simple mode-finding algorithm, analogous to the local linear approximation (LLA) of \citet{zou:li:2008}. 
 
Sections \ref{sec:n>p} and \ref{sec:p>n} illustrate applications of the approach in high-dimensional regression problems, including those where $p>n$, by placing local shrinkage priors on certain linear combinations of the $\beta_j$'s.  These linear combinations are given by the right-singular vectors of the design matrix.  Our approach therefore builds upon the work of \citet{frank:friedman:1993}, \citet{clydeJASA1996}, \citet{denison:george:2000}, \citet{west03}, and \citet{maruyama:george:2008}.  These authors provide a unified framework for ridge regression (RR), principal-component regression (PCR), partial least-squares (PLS), the $g$-prior, and generalized $g$-priors.  We generalize this framework still further by combining it with the idea of using local-shrinkage priors derived from L\'evy processes.

\section{Local shrinkage priors}
\label{sec:globallocal}


The class of joint priors $p(\bbeta)$ based on exchangeable normal variance mixtures (\ref{global-local1}--\ref{global-local3}) includes widely known forms such as the $t$ and the double-exponential, along with some of the following, more recent proposals.

\begin{description}
\item[Normal/Jeffreys,] where $\p(\beta_j) \propto |\beta_j|^{-1}$ \citep{figueiredo:2003, bae:mallick:2004}.  It arises from placing Jeffreys' prior upon each local variance: $\p(\lambda_i^2) \propto 1/\lambda_i^2$.
\item[Normal/exponential-gamma,] where $\lambda_j^2 \sim \Ex(r)$, and where there is a second-level $\mbox{Ga}(c,1)$ prior for the exponential rate parameter $r$ \citep{griffin:brown:2005}.  Marginally, this gives $\p(\lambda_i^2) \propto \left(  1+ \lambda_i^2 \right)^{-(c-1)}$.
\item[Normal/gamma and normal/inverse-Gaussian,] where the local variances receive gamma or inverse-Gaussian mixing densities \citep{caron:doucet:2008, griffin:brown:2010}.
\item[Horseshoe prior,] a special case of a normal/inverted-beta class, where $\lambda_i^2 \sim \IB(a,b)$ has an inverted-beta distribution \citep{Carvalho:Polson:Scott:2008a,polson:scott:2009a}.
\item[Generalized double-Pareto,] which has a Laplace-like spike at zero and polynomial tails \citep{dunson:armagan:lee:2010}.
\end{description}
Full posterior inference under these priors can be viewed as a Bayesian analogue of penalized-likelihood estimation.  For a more extensive bibliography, see \citet{Polson:Scott:2010a}.

These priors are typically used when $\bbeta$ is expected to be sparse.  A natural question is: why should Bayesians consider such an approach to a sparse problem, when these local-shrinkage priors do not explicitly allow for the possibility that some of the $\beta_j$'s are zero with positive prior probability?  At least three reasons suggest themselves.

First, suppose that one proceeds in the traditional Bayesian way, by averaging over different submodels in 
proportion to their posterior probabilities.  These model-averaged coefficients will be nonzero with probability 
1 under the sampling distribution for $\by$, regardless of $\bbeta$, and hence may be practically 
indistinguishable from the posterior mean of $\bbeta$ a carefully chosen shrinkage prior. 

Second, many Bayesians oppose testing point null hypotheses, and would rather shrink than select, on the grounds that point nulls are unrealistic.  Sparse shrinkage priors offer a compromise.  They discount the possibility that $\beta_j = 0$, yet they sift signals from noise more aggressively than a traditional elliptical prior.

Finally, the pure-shrinkage answer can offer computational gains over Bayesian model averaging.  For a normal linear model with conjugate priors, the difference may be small.  But for cases where marginal likelihoods of different regression hypotheses cannot be computed in closed form, the difference may be substantial, and the shrinkage approach can be used to approximate the model-averaged solution.

To illustrate this third argument, we simulated data from a probit model with $p = 25$ and $n=500$:
\begin{eqnarray*}
y_i &=& 1_{z_i > 0} \, \, \, \mbox{for}  \, \, \, i = 1, \ldots, n  \\
\boldsymbol{z} &\sim& \N(X\bbeta, I) \, ,
\end{eqnarray*}
where $\bbeta$ contained 20 zeros along with 5 nonzero entries, all equal to $\sqrt{5}$---a so-called ``$r$-spike signal'' with $r=5$ and $\Vert \bbeta \Vert^2 = p$.  The rows of $X$ were simulated from a multivariate normal distribution whose covariance matrix was drawn from an inverse-Wishart distribution, centered at $I_p$ and with $p+2$ degrees of freedom.

We simulated 100 data sets from this model, and compared four approaches for estimating $\bbeta$ using the probit link function: (1) maximum likelihood, using the \verb|glm| function in R; (2) lasso-CT, using the lasso penalty and choosing $\nu = \sqrt{2 \log p}$ as in \citet{candes:tao:2007}; (3) lasso-CV, with $\nu$ chosen by cross-validation; and (4) HS, the horseshoe posterior-mean estimator \citep{Carvalho:Polson:Scott:2008a}, a recent example of a pure-shrinkage approach designed to estimate sparse signals.  We measured accuracy in estimating $\bbeta$ by squared-error loss.  Table \ref{tab:probitexample} shows the median and mean sum of squared errors realized over the 100 simulations.  
The pure-shrinkage Bayesian model outperformed the alternatives by a wide margin.

Bayesian model averaging would be difficult here: the marginal likelihood for a given submodel cannot be computed in closed form, even assuming a conditionally conjugate prior for $\bbeta$.  Either high-dimensional numerical integration or a Laplace approximation must be used instead.  By contrast, a pure-shrinkage model is no harder to fit for binary data than it is for continuous data, using the simple trick of data augmentation.

\begin{table}
\begin{center}
\caption{\label{tab:probitexample} Median and mean sum of squared errors in reconstructing the probit $r$-spike signal in 100 simulated data sets.}
\vspace{1pc}
\begin{tabular}{r r r r r}
& MLE & Lasso-CT & Lasso-CV & HS \\
\hline
Median SSE & 19.0 & 15.3 & 12.3 & 0.7 \\
Mean SSE & 68.6 & 15.4 & 11.7  & 1.6 
\end{tabular}
\end{center}
\end{table}


This example motivates the question of how one should choose a prior $\pi(\lambda_j^2)$, 
or equivalently a penalty function, since different choices can 
lead to large differences in performance.  The oracle property provides a unifying framework for evaluating procedures under a classical framework; many different criteria have been proposed for accomplishing the same goal under a Bayesian framework.  To our knowledge only the lasso has been studied extensively under both paradigms.

One interesting question is: how can we translate between the Bayesian and penalized-likelihood formulations?  In the following section, we use the theory of L\'evy processes to establish a series of three (successively more general) characterizations of shrinkage priors and their relationship with penalty functions.

\section{Priors and penalties from L\'evy processes}
\label{sec:levyprocess}

\subsection{Normal variance mixtures and subordinated Brownian motion}

Our goal is to provide a framework in which important features of a prior for a high-dimensional location vector $\bbeta$ can be studied in terms of the L\'evy measure $\mu(dx)$ of some L\'evy process.  This perspective gives applied modelers a large toolbox for constructing prior distributions or penalty functions with specific desired properties.

This approach is most readily introduced via the special case of (\ref{global-local1})--(\ref{global-local3}) studied by \citet{caron:doucet:2008} and \citet{griffin:brown:2010}, where the normal--gamma prior for $\bbeta$ is seen to be the finite-dimensional marginal distribution of a variance-gamma process.

Let $T(s)$ be a standard gamma process having marginal distribution $T(s) \sim \Ga(s, 1)$ at time $s > 0$.  Because the gamma distribution is self-similar, for any value of $p$
$$
T(\nu) \stackrel{D}{=} \sum_{j=1}^p \lambda^2_j 
$$
if $(\lambda_j^2 \mid \nu) \stackrel{iid}{\sim} \Ga(\nu/p, 1)$, or equivalently if the $\lambda_j^2$'s are identified with the increments of $T$:
$$
\lambda_j^2 \stackrel{D}{=} T\left( \nu \cdot \frac{j}{p} \right) - T \left(\nu \cdot \frac{j-1}{p} \right) \, .
$$
As $p$ diverges, one may identify each local variance $\lambda_j^2$ with precisely one of the countable jumps in the sample path of the gamma process.  A tangential but interesting fact is that, if we were to normalize the $\lambda_j^2$'s by their sum $T(\nu)$, we would obtain the joint distribution for the weights in a Dirichlet-process mixture model \citep{kingman:1975}.

The gamma process is just one example of a subordinator, or a one-dimensional L\'evy process that is nondecreasing with probability 1.  If $T(s)$ is a subordinator and $W(s)$ is a standard Wiener process, then the L\'evy process $Z(s) = W\{T(s)\}$ is an example of subordinated Brownian motion observed on a random irregular time scale, a construction first explored by S.~Bochner in the 1950's. The increments of $T$ yield the local variances $\{ \lambda_j^2 \}$, while the increments of $Z$ give us the regression coefficients $\{ \beta_j\}$.  When $T$ is a gamma subordinator, $Z$ is called a variance-gamma process.

Subordinated Brownian motion is the natural infinite-dimensional generalization of a normal variance mixture.  Specifying the subordinator is equivalent to specifying the mixing measure $p(\lambda_j^2)$.

In this way, one may define a joint distribution for $\bbeta$ by way of a single quantity: the marginal distribution of a subordinator $T$ at time $s = \nu$.  One may generate other joint distributions for $\bbeta$ via the same device of slicing up a subordinator into its increments, and identifying these increments with the variances $\{\lambda_j^2\}$ in a conditionally normal joint distribution for $\bbeta$.  If, for example, $T(\nu)$ is inverse-Gaussian, then each $\beta_j$ will have a normal/inverse-Gaussian distribution \citep[see, e.g.,][]{barndorffnielsen:1997a}.

An important feature of subordinators is that they are infinitely divisible.  This ensures that our construction remains sensible even in the infinite-dimensional limit.   For example, suppose that we identify the local variances $\lambda_j^2$ of $p$ different $\beta_j$'s with the increments of $T$, a subordinator, observed on a regular grid.  This $p$-variate random variable can then be described \textit{a priori} in terms of the behavior of a single random variable $T$, which specifies an easily interpretable aggregate feature of the $\bbeta$ sequence---namely, the sum of the local variances.  If we were then to consider $2p$ $\beta_j$'s instead, but wished to retain the same aggregate features of the (now longer) $\bbeta$ sequence, we must merely slice up the increments of the original subordinator on a finer grid.

Self-similarity is a more restrictive but very appealing property.  It will ensure that, as $p$ grows and we divide the subordinator into arbitrarily fine increments, the probabilistic structure of the local precisions remains the same---a useful fact if one wishes to study a procedure's asymptotic properties.  For an extensive discussion and further bibliography of asymptotic theory regarding L\'evy processes, see \citet{ait-sahalia:jacod:2009a}.

\subsection{Penalty functions and subordinators}

Not all interesting penalties can be easily interpreted in the same way as the normal--gamma.  For example, the lasso corresponds to an exponential mixing distribution for $\lambda_j^2$.  Yet a sum of exponentials is not itself exponential, making it difficult to interpret the lasso prior as the increments of subordinated Brownian motion.

Luckily the theory of subordinators can be used in a slightly different way to obtain an alternative characterization of priors and penalty functions.  Stated informally: all totally monotone penalty functions that vanish at zero correspond to priors that can be represented in terms of a subordinator.  For certain penalties, this subordinator naturally corresponds to the precision, rather than the variance, of a conditionally normal prior.  We present this construction in the following theorem, which provides a rich source of new penalty functions with explicit Bayesian formulations as mixtures of familiar distributions.  Throughout the following discussion, we let $t$ denote a dummy argument involving $\beta_j$.

\begin{theorem}
\label{the:subordinatorpenalty}
Let $\psi(t)$, $t>0$, be a nonnegative-real-valued, totally monotone function such that $\lim_{t \to 0} \psi(t) = 0$. 
\begin{description}
\item[Part A:] Suppose that these conditions are met for $t \equiv f(\beta_j)$.  Then the prior distribution $p(\beta_j \mid s) \propto \exp\{ -s \psi[f(\beta_j)] \}$, where $s>0$,  is the moment-generating function of a subordinator $T(s)$, evaluated at $f(\beta_j)$, whose L\'evy measure satisfies
\begin{equation}
\label{eqn:levyrepsubordinator}
\psi(t) = \int_0^\infty \{1 - \exp(-t x) \} \ \mu(\mbox{d} x) \, .
\end{equation}

\item[Part B:] Suppose that these conditions are met for $t \equiv \beta_j^{2}/2$.  Then $p(\beta_j \mid s) \propto \exp\{ -s \psi(\beta_j^{2}/2) \}$, where $s>0$, is a mixture of normals given by
\begin{equation}
p(\beta_j \mid s) \propto \int_0^{\infty} \N \big( \beta_j \mid 0, T^{-1} \big) \ T^{-1/2} p(T) \ d T \, ,
\end{equation}
where $p(T)$ is the density of the subordinator $T$, observed at time $s$, whose L\'evy measure $\mu(dx)$ satisfies (\ref{eqn:levyrepsubordinator}).
\end{description}
\end{theorem}

As an example, consider the bridge estimator, for which $\log p(\beta_j) = -\nu |\beta_j|^{\alpha}$.  Write this instead as $-\nu (\beta_j^2/2)^{\alpha/2}$, in which case the conditions of Theorem \ref{the:subordinatorpenalty} are met for $\alpha \in (0, 2]$.  The resulting normal mixture is easily recognized as the moment-generating function, evaluated at $t = \beta_j^2/2$, of a positive alpha-stable subordinator $T$ with stability index $\alpha/2$, observed at time $s = \nu$. This provides a very simple proof of the fact the exponential-power priors are normal mixtures \citep{west:1987}.

The special case of the lasso ($\alpha = 1$) leads to a Stable$(1/2)$ law for $T$.  This is equivalent to an inverse-Gaussian representation of the lasso prior on the precision scale:
\begin{eqnarray*}
e^{-\nu |\beta_j|} &=& \int_0^{\infty} e^{-T \beta_j^2/2} \ \frac{\nu}{\sqrt{2 \pi T^3}} e^{-\nu^2/(2T)} d T \\
\IN(0, \nu) &\stackrel{D}{=}& \sum_{j=1}^p \IN(0, \nu/p) \, .
\end{eqnarray*}
The inverse-Gaussian (IN) distribution, moreover, is self-similar: a sum of $p$ inverse-Gaussian precision terms is still inverse-Gaussian, an analytically convenient property which the lasso fails to exhibit on the $\lambda_j^2$ scale.  This provides an alternative to the lasso's well-known characterization in terms of an exponential mixing distribution for $\lambda_j^2$.

Though we do not consider the point at length, Part B can be extended to the case where $t \equiv |\beta_j|^{b}$, $b \in (0, 2]$, subject to further mild regularity conditions on $\psi$.  The prior $p(\beta_j)$ will be a mixture of exponential-power distributions---itself a mixture of normals, in which case the law of iterated expectation will be enough to establish the result.

We can also consider a mixture or Rao-Blackwellized penalty function as follows.  Suppose we define
$$
g(\bbeta) = - \log  \E \Big[  C_\nu \exp \Big\{ - \nu \sum_{j=1}^p \psi(t_j) \Big\} \Big] \, ,
$$
where the expectation is under a prior $p(\nu)$, and where $C_\nu$ is the normalization constant in $p(\beta \mid \nu)$.  Suppose that $\psi(t)$ satisfies the conditions of the previous theorem and that the prior for $\nu$ can be described in the same way by a subordinator $T(s)$ with L\'evy measure $\mu(d x)$.  Then since $T$ is a subordinator, its moment-generating function is
\begin{eqnarray*}
M_s(t) &=& \E \{ \exp(- t T(s)) \}  =  \exp\{ -s \chi(t) \} \\
\chi(t) &=& \int_0^\infty \{1 - \exp(t x) \} \mu(d x) \, ,
\end{eqnarray*}
To compute the mixture penalty function, simply evaluate this moment-generating function for $T(1)$ at $t = \sum_{i=1}^p \psi(t_j)$ to give
$$
g(\bbeta) = \chi \left\{ \sum_{i=1}^p \psi(\beta_j^2) \right\} \, ,
$$
where we have absorbed a factor of $C_\nu^{-1}$ into the implicit prior for $\nu$, to cancel with the normalization constant from $p(\beta \mid \nu)$.

Consider the example of bridge estimation with an alpha-stable prior for the regularization parameter.  Specifically, let $\log p(\beta_j \mid \nu) = -\nu |\beta_j|$, and let $T$ be an $\alpha$-stable subordinator $T_{\alpha}(s)$, $0 < \alpha < 1$, observed at time $s=1$.  Then $\psi(t) = \sqrt{t^2}$, and $\chi(t) = |t|^{\alpha}$.  Therefore the mixture penalty function is
$$
\chi \left\{ \sum_{i=1}^p \psi(\beta_j^2) \right\} = \left( \sum_{i=1}^p |\beta_j| \right)^{\alpha} \, ,
$$
with no nuisance parameters left to estimate.

\subsection{Nonlinear time changes and further examples}
\label{subsec:nonlineartimechanges}

An even more general approach for building priors from time-changed Brownian motion is to specify the following:
\begin{enumerate}
\item a self-similar random variable $z \stackrel{D}{=} \sum z_j$.
\item a transformation $u$ mapping $z_j$ to the positive reals.  Typical examples are the identity, inverse, and log.
\item Brownian motion observed at random time increments $\delta_j = u(z_j)$.
\end{enumerate}

This approach encompasses many other examples of time-changed Brownian motion not previously studied in the presence of sparsity.  These examples collectively speak to the power and generality of the approach considered here.  For example, \citet{bn:shephard:2001a} study the class of normal/modified-stable processes, where the mixing distribution is based on exponential and power tempering (or tilting) of a positive $\alpha$-stable subordinator.  Another interesting generalisation is the Normal-Lamperti distribution with mixing density
$$
p(\lambda_j^2) = \frac{\sin(\pi \alpha)}{\pi} \frac{(\lambda_j^2)^{\alpha-1}} {(\lambda_j^2)^{2\alpha} + 2(\lambda_j^2)^{\alpha} cos(\pi \alpha) + 1} \, , \quad \lambda_j^2 > 0 \, .
$$

The transformation $u$ accommodates cases where the mixing distribution $p(\lambda_j^2)$ is not obviously self-similar.  The horseshoe prior of \citet{Carvalho:Polson:Scott:2008a} provides an example.  In the usual hierarchical representation of this prior, one specifies a standard half-Cauchy distribution for the local scales: $\lambda_i \sim \C^+(0, 1)$.  This corresponds to
$$
p(\lambda_i^2) \propto (\lambda_i^2)^{-1/2} (1 + \lambda_i^2)^{-1} \, ,
$$
an inverted-beta (or beta-prime) distribution denoted $\IB(1/2, 1/2)$.  This generalizes to the wider class of normal/inverted-beta mixtures \citep{polson:scott:2009a}, where $\lambda_i^2 \sim \IB(a,b)$.  These mixtures satisfy the weaker property of being self-decomposable: if $\lambda_i^2 \sim \IB(a,b)$, then for every $0 < c < 1$, there exists a random variable $\epsilon_c$ independent of $\lambda_i^2$ such that $\lambda_i^2 = c \lambda_i^2 + \epsilon_c$ in distribution.

We omit the proof of the fact that the inverted-beta distribution is self-decomposable; see Example 3.1 in \citet{bondesson:1990}.  The consequence of this fact is that the horseshoe prior can be represented directly as subordinated Brownian motion.  The proof is not constructive, however, as the subordinator itself is not available in closed form.  The difficulty becomes plain upon inspecting the characteristic function of an inverted-beta distribution:
$$
\phi(t) = \frac{\Gamma(a+b)}{\Gamma(b)} \ U(a, 1-b, -it) \, ,
$$
where $U(x,y,x)$ is a Kummer function of the second kind.  A characteristic function of this form makes it very difficult to compute the distribution of sums of inverted-beta random variables.

Representing the horseshoe prior in terms of the increments of a self-similar L\'evy process can be done straightforwardly, however, on the log-variance scale, just as a self-similar representation of the lasso model can be found on the precision scale.

Suppose $\lambda_i^2 \sim \IB(a,b)$.  Then
$$
\lambda_i^2 \stackrel{D}{=} \frac{\kappa_i}{1-\kappa_i} \, ,
$$
where $\kappa_i \sim \Be(a,b)$.  Following \citet{fisher:1935}, if $z_i = \log \{\kappa_i/(1-\kappa_i)\}$, then
$$
p(z_i) = \frac{1}{\mbox{B}(a,b)} \frac{(e^{z_i})^a}{(1 + e^{z_i})^{a+b}} \, ,
$$
where $\mbox{B}(a,b)$ is the Beta function.  More generally we may assume that $z_i \sim \Z(a,b,\mu,\sigma)$, a $z$-distribution with density
$$
p(z_i) = \frac{2\pi}{\sigma \mbox{B}(a,b)} \frac{ [\exp\{ (z_i-\mu)/\sigma  \} ]^a }  {[1+\exp\{ (z_i-\mu)/\sigma  \} ]^{a+b}} \, 
$$
and characteristic function
\begin{equation}
\label{eqn:zcharfunc}
\phi(t) = \frac{\mbox{B} \left( a + \frac{i \sigma t}{2\pi}, b - \frac{i \sigma t}{2\pi}  \right) }{ \mbox{B}(a,b)} \ \exp(i\mu t)
\end{equation}
for $a > 0$, $b > 0$, $\sigma > 0$, $\mu \in \mathbb{R}$.

The $z$ distribution can then be recognized as the special case the generalized-$z$ (GZ) distribution, which has characteristic function
$$
\phi(t) = \left\{ \frac{\mbox{B} \left( a + \frac{i \sigma t}{2\pi}, b - \frac{i \sigma t}{2\pi}  \right) }{ \mbox{B}(a,b)} \right\}^{2\delta} \ \exp(i\mu t)
$$
for $\delta > 0$ \citep{grigelionis:2001}.  This distribution has parameters $(a,b,\mu,\sigma, \delta)$ and can also be characterized by its L\'evy triple $\{A, 0, \mu(x) dx\}$, where
\begin{equation}
\label{eqn:brownianpartZ}
A = \frac{\sigma \delta}{\pi} \int_0^{2\pi/\sigma} \frac{ e^{-bx} - e^{-ax} } {1 - e^{-x}} \dd x + \mu \, ,
\end{equation}
and
$$
\mu(x) = 
\left\{
\begin{array}{l l}
\frac{2\delta \exp \left\{ \frac{ 2\pi b x}{\sigma} \right \} }{x \left\{ 1 - \exp \left( \frac{2\pi x}{\sigma} \right) \right\}} \, , & \mbox{if $x>0$} \\ \\
\frac{2\delta \exp \left\{ \frac{ 2\pi a x}{\sigma} \right \} }{|x| \left\{ 1 - \exp \left( \frac{2\pi x}{\sigma} \right) \right\}} \, , & \mbox{if $x<0$} \, .
\end{array}
\right. 
$$

The characteristic function of a generalized-$z$ distribution makes its self-similarity plain: if $z_i \stackrel{iid}{\sim} \GZ(a,b,\mu/p,\sigma,1/2p)$, then
$$
\sum_{i=1}^p z_i \stackrel{D}{=} z \, ,
$$
where $z \sim \Z(a,b,\mu,\sigma)$.  We thus have a self-similar representation, on the log-variance scale, of the normal/inverted-beta class.

This result is of limited use except in special cases where the density of the generalized-$z$ increments is known, which will not hold in general.  Luckily the horseshoe prior, where $a=b=1/2$, corresponds to just such a special case---as do all symmetric cases where $\kappa \sim \Be(a,1-a)$ and $\lambda_i^2 = \kappa/(1-\kappa)$.

To see this, let $z \sim \Z(a, 1-a, \mu, \sigma)$ for $a \in (0,1)$.  Then standard manipulations of the characteristic function (\ref{eqn:zcharfunc}) give
$$
\phi(t) = \frac{\cos(c/2)}{ \cosh \big( \frac{\sigma t - ic} {2} \big) } \ \exp(i\mu t) \, ,
$$
where $c = \pi(2a-1)$.  This is recognizable as the characteristic function of a Meixner process, $z \sim \Meix(\sigma, c, 1/2, \mu)$ \citep{grigelionis:1999}.  The density and L\'evy measure of a Meixner random variable are
\begin{eqnarray}
p(z) &=& \frac{2\cos(c/2)}{\sigma \pi} \exp \left\{  \frac{c(z-\mu}{\sigma} \right\} \left| \Gamma \left( \frac{1}{2} + \frac{i (z - \mu)}{\sigma} \right) \right|^2 \\
\mu(dx) &=& \frac{\exp(cx/\sigma)}{2 x \sinh(\pi x/\sigma)} \dd x \, .
\end{eqnarray}
For the horseshoe prior, $a = 1-a$ and therefore $c=0$.

A Meixner process is self-similar: if $z_i \sim \Meix\{a,c,1/(2p), \mu/p \}$, then
$$
\sum_{i=1}^p z_i \stackrel{D}{=} z \sim \Meix(a,c,1/2, \mu) \, .
$$
When $a=1$ and $\mu=0$, then the random variable $T \stackrel{D}{=} e^z$ will have an $\IB(a,1-a)$ distribution, as required.  Therefore, the most intuitive way of passing to a limit under the horseshoe prior is to continue dividing the random variable $T$, on the log variance scale, into arbitrarily many self-similar increments.

Interestingly, both the $z$-distribution and the Meixner can themselves be represented as mixtures of normals.  The mixing distribution for the $z$ is an infinite convolution of exponentials, a potentially interesting generalization of the lasso model \citep{bn:kent:sorensen:1982}.  For the mixing distribution of the Meixner, see \citet{madan:yor:2006a}.

\section{The general L\'evy-process case}
\label{sec:generallevycase}

We have encountered two ways in a subordinator can be used to generate joint distributions for $\bbeta$, or equivalently penalty functions:
\begin{enumerate}
\item by subordinating Brownian motion to $T(s)$, leading to a L\'evy process $Z(s)$ whose increments are identified with the components of $\beta_j$.
\item by using the subordinator's Laplace exponent $\nu \psi(t)$ as a penalty function, which sometimes leads to a tractable mixture representation for the corresponding prior $p(\beta_j \mid \nu) \propto \exp \{-\nu \psi(t) \}$.
\end{enumerate}
In general Bayesians have focused on the finite-dimensional analogue of the first approach, while frequentists have focused on the second approach, although many authors have focused on explicit translations of a classical estimator into a Bayesian model \citep[e.g.][]{park:casella:2008,hans:2008}.

An encompassing formulation involving L\'evy processes is available.  This is most easily understood in the case of an orthogonal design matrix $X$, in which case we define $\tilde{\by} = X'\by$. Let $\Delta = \nu p^{-1}$, and let
$$
\beta_j \stackrel{D}{=} Z(j\Delta) - Z ([j-1]\Delta)
$$
for some arbitrary L\'evy process $Z(s)$ having L\'evy measure $\mu(dx)$, assumed to be defined over the interval $[0, \nu]$.  Then upon observing $\tilde{\by} = (\tilde{y}_1, \ldots, \tilde{y}_p)$ with $\tilde{y}_j \sim \N(\beta_j, \sigma^2)$, identify $\tilde{\by}$ with the increments of the interlacing process $Y(s) = Z(s) + \sigma_p W(s)$:
$$
\tilde{y}_j \stackrel{d}{=} Y(j\Delta) - Y([j-1] \Delta) \, .
$$
The observations are themselves the increments a L\'evy process: a superposition of signals or jumps identified with $Z(s)$, and noise identified with a scaled Wiener process $W(s)$.

The Bayesian local-shrinkage framework of Equations (\ref{global-local1})--(\ref{global-local3}) is to specify the distribution of the increment $\delta = Z(j\Delta) - Z ([j-1]\Delta)$ as a Gaussian mixture.  In general the corresponding L\'evy measure will not be known.  Unless the mixing distribution belongs to some convolution-closed family (such as the gamma or inverse-Gaussian), we will not know the distribution of increments at other ``time scales,'' and asymptotic analysis may be difficult.

The above construction says, in effect, that one can proceed by specifying the L\'evy measure directly, with the two subordinator-based approaches being intermediate cases.  Indeed, by the L\'evy-Khinchine theorem, any model that preserves the conditional-independence property of the $\beta_j$'s will fall into this framework, since any stationary c\`adl\`ag process with independent increments is completely characterized by its L\'evy measure.

By casting the finite-dimensional problem in terms of the marginal distributions of a suitable infinite-dimensional problem, the L\'evy process view provides an intuitive framework for asymptotic calculations.  Such analysis can be done under one, or both, of two assumptions: that we observe the process longer, or that we observe it on an ever finer grid.  Each scenario corresponds quite naturally to a different assumption about how the signal-to-noise ratio behaves asymptotically.

Finally, it is possible to generalize these methods still further, following along the lines of the nonparametric function-estimation strategy proposed by \citet{Wolp:Clyd:Tu:2010}.  These authors consider priors for kernel weights based on a stochastic integral of a generator function with respect to a random measure, which allows for the incorporation of spatial marks, periodicities, and further covariates into the prior \citep[see also][]{clyde:wolpert:2011a}.  Since our interest is in priors for $\bbeta$ that maintain exhangeability among the regression coefficients and thus correspond to traditional penalty functions, we do not pursue this approach here.

\section{The statistical relevance of the L\'evy-process view}

\label{sec:statisticalrel}

\subsection{L\'evy processes and the two-groups model}

We now describe, in a more precise way, the result mentioned in the introduction: that Bayesian variable selection and pure-shrinkage solutions like the lasso can both be viewed as special cases of the same encompassing framework.

The familiar discrete mixture or ``two-groups'' model specifies that each $\beta_j$ is either in or out of the model with some prior inclusion probability:
$$
\beta_j \sim w p(\beta_j) + (1-w) \delta_0 \, ,
$$
where $\delta_0$ is a Dirac measure.  This is the typical assumption used in Bayesian model selection, model averaging, and multiple testing \citep[c.f.][]{georgefoster2000,scottberger2007}.

The two-groups model arises as a special case of the L\'evy-process framework: namely, when $Z(s)$ has a finite L\'evy measure and is therefore a compound Poisson process.  Under this assumption,
$$
Y(s) = \sum_{i=1}^{N(s)} J_i + \sigma W(s) \, ,
$$
where $N(s)$ is a Poisson process with rate $\theta$ governing the number of jumps that occur by time $s$, and each $J_i$ is an independent draw from some jump distribution.

With probability 1, a compound Poisson process will have a finite number of jumps on any finite interval.  These jumps correspond to the nonzero signals in $\bbeta$; all other increments of $Z(s)$ will be zero.  The L\'evy density of $Z(s)$ describes the distribution of the signals, while the jump rate (which can be identified in terms of the total mass of the L\'evy density) describes their relative abundance in the cohort of $\beta_j$'s under consideration.

To illustrate the connection, suppose that $J_i \sim \N(0, \eta^2)$, and that we follow the previous line of reasoning by equating the regression coefficients $\beta_j$ with the increments of $Z(s)$ on a discrete grid of size $\Delta$.  Then with probability $w = 1- e^{-\theta \Delta}$, $\beta_j$ will correspond to an interval where at least one jump has occurred.  Moreover, each nonzero $\beta_j$ will arise from a normal distribution:
\begin{eqnarray*}
\beta_j &\sim& \N(0, \tau^2) \\
\tau^2 &=& \sum_{k=1}^{\infty} \frac{w_k^2 \eta^2}{w^2} \, ,
\end{eqnarray*}
where $w_k = (k!)^{-1} (\Delta \theta)^k \exp(-\Delta \theta)$ is the probability of seeing $k$ jumps. In essence, the missing $k=0$ term corresponds to the null hypothesis of no jumps, yielding $\beta_j = 0$.

The discrete-mixture prior is an example of a finite-activity process where the total L\'evy measure is finite.  But one could also use an infinite-activity process, where the L\'evy measure is merely sigma-finite.  This would mean that the underlying process had an infinite number of very tiny jumps---in other words, that no $\beta_j$'s are zero, but that most are of insignificant size compared to $\sigma$.  The pure-shrinkage (``one-group'') model and the two-groups model can therefore be subsumed into this single framework.

An interesting question is: how different are the one-group and two-group models, asymptotically (i.e.~as $p \to \infty$, and therefore $\Delta \to 0$)?  Observe that under the two-groups model where $Z(s)$ is a compound Poisson process with jump density $g$,
$$
P(|\beta_j| > \epsilon) = \Delta \theta \int_{\Omega(\epsilon)} g(x)  \dd x + o(\Delta) \, ,
$$
where $\Omega(\epsilon)$ = $(-\infty, \epsilon) \cup (\epsilon, \infty)$.  This decreases linearly in $\Delta$, at a rate governed by the jump activity $\theta$ of the Poisson process.

Meanwhile, if $Z(s)$ is instead a pure-jump L\'evy process with L\'evy measure $\mu(dx)$, then
$$
P(|\beta_j| > \epsilon) = \Delta \int_{\Omega(\epsilon)} \mu(dx) + o(\Delta) \, .
$$
Any L\'evy process necessarily assigns finite measure to the set $\Omega(\epsilon)$ for $\epsilon > 0$, so this probability also decreases linearly in $\Delta$.  In this sense, the class of priors derived from the increments of a L\'evy process encompasses those priors that can be made to asymptotically mimic the one-group model in terms of the measure they assign to $\Omega(\epsilon)$ for any $\epsilon > 0$.  An interesting comparison is with the work of \citet{berger:delampady:1987}, especially their discussion concerning the validity of approximating interval nulls by point nulls.

\subsection{A representation of the posterior mean}

Much of the research on penalized-likelihood estimation concerns methods for finding sparse posterior modes in high-dimensional regression problems.  Yet the posterior mean is the estimator that minimizes posterior expected loss under the squared-error loss function, and can lead to improved predictions compared to the posterior mode \citep[c.f.][]{efron:2009}.  It is therefore interesting to compare the behavior of the posterior mean estimator under different joint distributions for $\bbeta$.

We again consider the orthogonal-design case, or the exchangeable normal-means problem.  Recall the following result from \citet{pericchi:smith:1992}.  If $p(y-\beta)$ is a normal likelihood of known variance $\sigma^2$, $p(\beta)$ is the prior for $\beta$ (subject to some mild regularity conditions), and $m(y) = \int p(y - \beta) p(\beta) \ \dd \beta$ is the predictive density for $y$, then:
\begin{equation}
\label{pericchismith}
\E(\beta \mid y) = y + \sigma^2 \frac{\dd}{ \dd y} \ln m( y ) \, .
\end{equation}

This result is useful for the insight it gives about an estimator's behavior in situations where $y$ is very different from the prior mean.  In particular, it shows that ``Bayesian robustness'' may be achieved by choosing a prior for $\beta$ such that the derivative of the log predictive density is bounded as a function of $y$.  Models meeting the slightly stronger condition that $[E(\beta \mid y) - y] \to 0$ for large $|y|$ are said to have redescending score functions.  

We generalize this result as follows.

\begin{theorem}
\label{thm:masreliezextension}
Let $p(|y - \beta|)$ be a likelihood that is symmetric in $y - \beta$.  Let $\psi(t)$ be a penalty function satisfying the conditions of Theorem \ref{the:subordinatorpenalty} for $t = \beta^2 /2$, and for which the corresponding subordinator $T \equiv T(s)$, $s>0$, has a prior $p(T)$ satisfying $E\{T^{-1}\} < \infty$.  Define the following size-biased pseudo-density and corresponding marginals:
\begin{eqnarray*}
p^\star (T ) &=& \frac{ T^{-1} p (T)}{ E(T) } \\
p^\star ( \beta) &=& \int_0^\infty e^{ - T \beta^2 /2}  p^{\star}(T)  \ \dd T \\
m^\star ( y) &=& \int_0^{\infty} p(y - \beta) p^{\star}(\beta) \ \dd \beta \, .
\end{eqnarray*}
Then
\begin{equation}
E( \beta \mid y )  = E(T^{-1}) \frac{m^\star (y)}{m(y)} \frac{\partial}{\partial y} \ln m^\star ( y) \, .
\end{equation}
\end{theorem}

Special cases of this theorem have appeared repeatedly in the literature; c.f.~\citet{masreliez:1975}, \citet{polson:1991}, \citet{mitchell:1994}, \citet{Carvalho:Polson:Scott:2008a}, and \citet{griffin:brown:2010}.  These results have been used to characterize ``good'' mixing distributions $p(\lambda_j^2)$ in the traditional global-local shrinkage model.  The important insight is that the sparse signal-detection problem is essentially the same as the outlier-sensitivity problem, a classic topic of interest in robust Bayesian statistics.

Our result extends this long line of research to provide a more general expression for the posterior mean.  It uses the subordinator representation to characterize the posterior mean corresponding to any penalty function meeting the regularity conditions of Theorem \ref{the:subordinatorpenalty}.  The intuition is essentially that, whenever a prior is chosen such that $m^{\star}(y)$ has a small derivative in a large neighborhood of the origin, the posterior mean will strongly shrink small observations to $0$.  The result also directly describes an estimator's sensitivity to aberrant observations---that is, signals---in terms of the corresponding L\'evy measure, rather than the prior for the local-shrinkage parameter $\lambda_j^2$.

Extending the general approach to non-orthogonal designs is straightforward, but algebraically involved.  It follows closely the method of proof pursued by \citet{masreliez:1975} and \citet{griffin:brown:2010}.

\subsection{Finding sparse solutions via the posterior mode}


Using Theorem \ref{the:subordinatorpenalty}, we can also develop simple EM algorithms for estimating the posterior mode of $\beta$ under a wide variety of models.

First, one may express a wide variety of problems as mixtures of ridge regressions, following along the lines of \citet{caron:doucet:2008} and \citet{dunson:armagan:lee:2010}.  If we take $  f ( \beta_j ) = \frac{1}{2} \beta_j^2 $ as in the previous theorem, then a similar line of reasoning leads to an algorithm for finding the mode, rather than the mean.  Under the conditions of theorem 1, suppose we have
\begin{equation}
\label{eqn:LQAidentity1}
e^{-\nu \psi(\beta_j^2/2)} = \int_0^{\infty}  e^{-T_j \beta_j^2/2} p(T_j) \ d T_j \, .
\end{equation}
Given a set of augmentation variables $\{T_j\}$, the conditional log-posterior distribution becomes
$$
 l ( \beta ) = \sum_{i=1}^n l_i ( \beta ) -  \sum_{j=1}^p T_j \beta_j ^2/2  \, ,
$$
where $l_i$ is the log-likelihood contribution associated with observation $i$.  For a normal likelihood, this will be the log density of a normal posterior whose mode is the generalized ridge estimator
$$
\hat{\bbeta} = (X'X + \nu^2 \mathbf{T}) X' \by \, ,
$$
where $\mathbf{T} = \mbox{diag}(T_1, \ldots, T_p)$.

This provides the M step.  Moreover, since the complete-data log likelihood is linear in $T_j$, its expected value given a current estimate $\bbeta^{(g)}$ is
$$
Q(\bbeta) =  \sum_{i=1}^n l_i ( \beta ) -  \sum_{j=1}^p \E\big( T_j \mid \beta_j^{(g)} \big) \beta_j ^2/2  \, .
$$
This expectation can be computed by differentiating (\ref{eqn:LQAidentity1}) under the integral sign to give
$$
\E \big( T_j \mid \beta_j \big) = \frac{\psi'(\beta_j^2 / 2)}{|\beta_j|} \, .
$$
Plugging in the current estimate $\beta_j^{(g)}$ gives the E step.

A second algorithm motivated by Theorem \ref{the:subordinatorpenalty} generalizes the LLA approach of \citet{zou:li:2008}.  Suppose we take $ f ( \beta_j ) = | \beta_j | $.  Then if $\psi(|\beta_j|)$ meets the conditions of the theorem, it is the log-moment generating
function of a subordinator $T_j \equiv T(\nu)$, and
\begin{align*}
 \exp \{ -\nu \psi ( | \beta_j | ) \} & = \int_0^\infty e^{ - T_j | \beta_j | } 
 p( T_j ) d T_j  \\
 - \nu \psi ( | \beta_j | ) & = \log \int_0^\infty e^{ - T_j | \beta_j | }
 p( T_j ) d T_j \, .
\end{align*}
Recall that the time $\nu$ at which the subordinator is observed corresponds to the global regularization parameter, assumed to be given. Taking derivatives with respect to $ \beta_j $ inside the integral sign gives us the identity
$$
 sign ( \beta_j ) \cdot \nu \psi' ( | \beta_j | ) = E  ( T_j  \mid \beta_j  ) \, .
$$
Here the expectation is with respect to the conditional posterior
$$
p \left ( T_j \mid \beta_j \right ) \propto e^{ - T_j | \beta_j | } 
 p( T_j ) \, .
$$

Moreover, observe that the complete-data log-likelihood using $ T_j $ as an augmentation variable takes a simple form:
$$
 l ( \beta ) = \sum_{i=1}^n l_i ( \beta ) -  \sum_{j=1}^p T_j | \beta_j |  \, ,
$$
This expression in linear in $T_j$, which suggests a simple EM algorithm.  Suppose we have a current estimate $\beta_j^{(g)}$.  For the E-step, we take the conditional expectation of the likelihood $ l ( \beta ) $ 
with respect to $p( T_j | \beta_j ) $ to obtain the objective function
$$
Q(\bbeta) = \frac{1}{n}\sum_{i=1}^n l_i ( \beta ) - \sum_{j=1}^p 
 E \left ( T_j  \mid \beta_j^{(g)} \right ) | \beta_j |  \, ,
$$
where $ E \left ( T_j \mid \beta_j \right ) = sign ( \beta_j ) \cdot \nu \psi^\prime ( | \beta_j | )  $.  This is the usual convex optimization problem encountered in finding a lasso solution, meaning that the M-step can be solved painlessly, using standard methods.  Lasso is already in this form without the need for augmentation variables, but other models representable as mixtures of double-exponentials are just as simple to fit.

An illuminating comparison is with the local-linear-approximation algorithm (LLA) of \citet{zou:li:2008}, specifically Equations 2.7 and 2.10.  Specifically, our Theorem \ref{the:subordinatorpenalty} generalizes 2.10 to cases beyond Laplace transforms of double-exponentials, and provides a probabilistic interpretation for all penalty functions in the class by expressing the corresponding Bayesian scale-mixture models in terms of an underlying L\'evy measure.  This probabilistic interpretation also leads to the expressions for the posterior mean derived in the previous subsection.

\section{Regularized regression when $p < n$}
\label{sec:n>p}

\subsection{Connections among RR, PCR, PLS, and the $g$-prior}

Thusfar we have considered L\'evy processes for constructing high-dimensional joint prior distributions for regression coefficients in a manner than maintains the exchangeability of the $\beta_j$'s.  This nests the traditional local-shrinkage approach in (\ref{global-local1})--(\ref{global-local3}), considered by many authors.  We now consider the mroe general case where the object of inferential interest is not necessarily $\bbeta$, but a set of linear combinations thereof---an approach that will generalize more easily to the $p>n$ case.  In particular we specify priors in the coordinate system defined by the principal components of $X'X$, although in principle other linear combinations follow the same template.  This will illuminate connections among the work of \citet{frank:friedman:1993}, \citet{west03}, \citet{maruyama:george:2008}, and ours on L\'evy processes.

Let $X = UDW'$ represent the singular-value decomposition of the design matrix $X$.  If $n > p$, then $X$ is of full column rank, and $D = \mbox{diag}(d_1, \ldots, d_p)$ is a  diagonal matrix of nonzero singular values ordered $d_1 > \cdots > d_p$.  Both $U$ and $W$ are orthogonal matrices, of dimensions $n \times p$ and $p \times p$, respectively.  Moreover, $W$ is also the matrix of eigenvectors $\{w_j\}$ for the cross-product matrix $S = X'X$, with corresponding eigenvalues $d_j^2$.

The original regression relationship may be re-expressed in the orthogonalized space as $y = Z \balpha + \epsilon$, where $Z = UD$ and $\balpha = W' \bbeta$.  The ordinary least-squares (OLS) estimate for $\balpha$ is $\hat{\balpha} = (Z'Z)^{-1} Z' y = D^{-1} U' y $.

Following \citet{frank:friedman:1993}, the shrinkage structures for many common regularization approaches can be understood by expanding their solutions in the original coordinate system in terms of the eigenvectors $\{w_1, \ldots, w_p \}$ and the OLS coefficients $\hat{\balpha}$:
\begin{equation}
\label{eqn:shrinkage.general}
\hat{\bbeta}^M = \sum_{j=1}^p \kappa_{j}^M \hat{\alpha}_j w_j \, .
\end{equation}
Here $M$ denotes the method, and the $\kappa_j^M$'s are method-specific shrinkage weights that scale the OLS solution along each of the directions $w_j$.

Both ridge regression and principal-components regression use shrinkage weights that do not depend on the response values $\by$.  The ridge-regression solution is $\kappa_j^{RR} = d_j^2 / (\nu + d_j^2)$ for a fixed regularization parameter $\nu$, while the $K$-component PCR solution is
$$
\kappa_{jK}^{PCR} = 
\left\{
\begin{array}{l l}
1, & d_j^2 \geq d_K^2 \\
0, & d_j^2 < d_K^2
\end{array}
\right.
\, .
$$
The posterior mean under the $g$-prior also fits in this shrinkage structure; it corresponds to $\kappa_j^{g} = g/(1+g)$, thereby shrinking the solution vector along all eigen-directions by a common factor.

The shrinkage weights under partial least squares, on the other hand, depend nonlinearly upon the response values $\by$ through the OLS solution $\hat{\balpha}$.  Using the expressions in \citet{frank:friedman:1993}, for the $K$-component solution we have
$$
\kappa_{jK}^{PLS} = \sum_{k=1}^K \theta_k d_j^{2k} \, ,
$$
where $\btheta = \{\theta_1, \ldots, \theta_K\}'$ is equal to $W^{-1} \eta$, with
$$
\eta_k = \sum_{j=1}^p \hat{\alpha}_j^2 d_j^{2(k+1)} \; {\rm and} \; 
W_{kl} =  \sum_{j=1}^p \hat{\alpha}_j^2 d_j^{2(k+l+1)} \, .
$$

\subsection{A Bayesian interpretation}

These four procedures differ only in the way that they scale the OLS estimates for the regression parameter in the orthogonal coordinate system defined by $W$.  It is therefore natural to consider them as special cases of an encompassing local-shrinkage model along the lines of the previous sections. 

Begin with the $g$-prior, an explicitly Bayesian model wherein $\bbeta \sim \N\{0, \sigma^2 g (X'X)^{-1}\}$ \textit{a priori}, or equivalently $\balpha \sim \N(0, \sigma^2 g D^{-2})$.  This prior biases the direction of $\alpha$ along the axes of the principal-component coordinate system.

Ridge regression also has a well-known Bayesian interpretation as the posterior mean under the conjugate normal prior $\bbeta \sim \N(0, \sigma^2 \tau^2 I)$, where the global variance $\tau^2 = 1/\nu$.  This prior is agnostic with respect to the orientation of the regression vector, depending only upon its Euclidean norm.

These procedures, along with PCR, are all special cases of a more general prior:
\begin{equation}
\label{eqn:globallocalprior}
(\balpha \mid \sigma^2, \tau^2, \Lambda)  \sim \N(0, \sigma^2 \tau^2 \Lambda) \, ,
\end{equation}
where $\tau^2$ is a global variance component and $\Lambda = (\lambda_1^2, \ldots, \lambda_p^2)$ is a diagonal matrix of local variance components.  The posterior distribution of $\balpha$ under this prior is conditionally normal, with mean
$$
m_j = \kappa_j \hat{\alpha}_j =  \left( \frac{\tau^2\lambda_j^2 d_j^2}{1 + \tau^2 \lambda_j^2 d_j^2} \right) \hat{\alpha}_j \, ,
$$
with the $\alpha_j$'s being mutually independent given $\tau^2$, $\sigma^2$, and the data.

The classical $g$-prior therefore corresponds to $\tau^2 = g$ and $\lambda_j \equiv d_j^{-2}$.  Ridge regression corresponds to $\lambda_j^2 = 1$.  And PCR corresponds to
$$
\lambda_j^2 = 
\left\{
\begin{array}{l l}
\infty, & d_j^2 \geq d_K^2 \\
0, & d_j^2 < d_K^2
\end{array}
\right.
\, 
$$
for the $K$-component solution.

Rather than estimating $\balpha$ under fixed choices of the local variances $\lambda_j^2$, the natural fully Bayesian approach is to use the shrinkage weights
\begin{equation}
\label{eqn:BayesWeights}
\kappa_{j}^{FB} = \E_{(\lambda_j^2, \tau^2 \mid X, \by)} \left(  \frac{\tau^2\lambda_j^2 d_j^2}{1 + \tau^2 \lambda_j^2 d_j^2} \right) \, ,
\end{equation}
where the expectation is over the posterior distribution of local and global variance components.

Different choices for the priors $\p(\lambda_j^2)$ and $\p(\tau^2)$ can center the Bayesian model at 
different classical regularization approaches, while still allowing the data to dictate otherwise.  
Choosing $\p(\lambda^2_j)$ to concentrate near $1$, for example, will center the model near the classical 
ridge solution.  On the other hand, if $\lambda_j^2 \equiv d_j^{-2} v_j^2$, then choosing $\p(v_j^2)$ to concentrate near $1$ will center the model near the $g$-prior.  Placing a further prior on $\tau^2$ will replicate the mixtures of $g$-priors studied by \citet{liangpaulo07}.

Mixing over a further prior $\p(\Lambda)$, however, will lead to even more flexible mixtures of $g$-priors.  In particular, the classical $g$-prior prefers coefficient vectors that line up with the principal components, and further mixing over local variance components helps to robustify the model against this assumption.

Even the PCR solution can be chosen as an approximate centering model by selecting a prior $\p(\lambda_j^2)$ such that $\p(\kappa_j)$ concentrates simultaneously near $0$ and $1$.  For example, if $\tau^2 = 1$ and $\lambda_j^2$ follows an inverted-beta (or ``beta-prime'') distribution $\IB(1/2,1/2)$, then $\kappa_j$ will have a $\Be(1/2,1/2)$ prior, whose density function is unbounded both at 0 and at 1 as required.  Marginally this leads to a horseshoe prior for $\alpha_j$ \citep{Carvalho:Polson:Scott:2008a}.

Partial least squares, on the other hand, cannot be interpreted in this framework.  To see this, observe that the shrinkage weights are identified with the prior variance components via $\kappa_j = \tau^2\lambda_j^2 d_j^2 /(1 + \tau^2 \lambda_j^2 d_j^2)$.  Under PLS, some of the shrinkage weights $\kappa_{jK}^{PLS}$ may be larger than $1$.  Such weights cannot arise from a valid (non-negative) configuration of $\lambda_j^2$'s and $\tau^2$.  Therefore, PLS cannot be the optimal solution under any prior expressible as a global-local scale mixture of normals. 

\subsection{When should the full Bayesian model work better?  Some intuition and examples}

Ridge regression, PCR, and PLS are all operationally similar.  They bias the coefficient vector away from directions in which the predictors have low sampling variance---or equivalently, away from the ``least important'' principal components of $X$.  This leads to a favorable bias-variance tradeoff in the performance of the resulting estimator.  The $g$-prior and mixtures of $g$-priors, on the other hand, shrink along all eigen-directions equally, and usually not by very much.

Neither of these approaches need work well.  When the underlying regression signal is ``eigen-sparse''---that is, when only some of the linear combinations of $\beta_j$'s given by $W$ are meaningful for predicting $\by$---then one should shrink different components of $\hat{\balpha}$ by different amounts.  This makes the $g$-prior inappropriate.

Yet as many previous authors have noted, there is no logical reason that $\by$ cannot be strongly associated with the low-variance principal components of $X$.  Ridge regression and PCR will both do poorly in these situations: RR will necessarily shrink more along low-variance directions, while PCR must include all the higher-variance directions ($j<K)$ in order to include a lower-variance one ($K$).

The intuition behind the fully Bayes model of (\ref{eqn:globallocalprior}) is that the shrinkage weights $\kappa_j$ should indeed be unequal, but that they can be learned from the data, and need not be monotonic in $d_j^2$.  The fully Bayes shrinkage weights, moreover, will depend not merely on $X$.  They will also depend nonlinearly upon $\by$, and upon each other through their mutual dependence upon $\tau^2$.

Consider three illustrative examples.  Although there are many  options to explore using the results of previous sections, in all cases we have assumed for the sake of illustration that $\tau^2 \sim \IB(1/2,1/2)$ and that $\lambda_j^2 \sim \IB(1/2,1/2)$, thereby specifying a geometric-Meixner-process prior for $\balpha$ (see~Section \ref{subsec:nonlineartimechanges}).

First, we analyzed the data from \citet{fearn:1983}, consisting of 24 samples of ground wheat.  The response variable is the protein concentration in the wheat, while the predictors (L1--L6) are  measurements of the samples' reflection of NIR radiation ($R$), measured at six different wavelengths between 1680 and 2310 nanometers.  The predictors are referred to as ``log values'', since they are measured on a $\log(1/R)$ scale.  The goal is to find a linear combination of log values that predicts protein concentration.  Both the response and the predictors were centered and rescaled to have variance 1.

The log values are highly multi-collinear, with the smallest pairwise correlation being $0.925$.  Despite the fact that ridge regression is intended for just these multi-collinear situations, here it performs quite poorly.   As \citet{fearn:1983} explains, this happens because the first principal component places nearly equal weight on all six log values (see Table \ref{tab:fearndata}).  The variation described by this component---essentially the sample average of the log values---is due mainly to differences in particle size.  It carries little information about protein content, and yet is prefentially treated as the ``most important'' predictor by the ridge estimator.  Contrasting log values are associated with ``less important'' principal components, and yet these contrasts---mostly the second, third, and fourth---are far more useful for predicting protein concentration.  Ridge regression shrinks these components more aggressively than the other methods.  Also observe the large amount of uncertainty surrounding the higher-order shrinkage factors.

\begin{table}
\begin{center}
\begin{footnotesize}
\caption{\label{tab:fearndata} The six principal component variances and loadings for the wheat protein-concentration data.}
\vspace{1pc}
\begin{tabular}{r r r r r r r}
     &   PC1 & PC2 & PC3 & PC4 & PC5 & PC6\\
L1 & 0.411 & 0.213 & 0.265 & -0.353 & 0.422 & 0.642\\
L2 & 0.410 & 0.342 & -0.446 & -0.079 & 0.465 & -0.542\\
L3 & 0.411 & 0.266 & -0.367 & -0.209 & -0.743 & 0.173\\
L4 & 0.411 & -0.028 & 0.731 & -0.127 & -0.221 & -0.481\\
L5 & 0.396 & -0.874 & -0.242 & -0.126 & 0.067 & 0.023\\
L6 & 0.411 & 0.05 & 0.05 & 0.891 & 0.013 & 0.182\\
\hline
     Variance & 5.868 & 0.101 & 0.019 & 0.012 & $< 0.001$ & $<0.001$
\end{tabular}
\end{footnotesize}
\end{center}
\end{table}

Second, we analyzed data on the softening temperature ($y$) of $n=99$ ash samples originating from different biological sources.  The predictor matrix comprises $p=16$ observed mass concentrations for the ash samples' constituent molecules.  The measurements are highly multi-collinear, with the eigenvalues of the correlation matrix for $X$ spanning 10 orders of magnitude.  The data are available in the R package \verb|chemometrics|, and have been centered and scaled.

Finally, we analyzed synthetic data where $X$ corresponds to a factor model.  That is, each row $x_i'$ satisfies
$$
x_i = B f_i + \xi_i \, ,
$$
where the loadings matrix B is $p \times k$, $f_i \sim \N(0, I)$ is $k \times 1$, $\xi_i \sim \N(0, \psi I)$ is $p \times 1$, and $k < p$.  The predictors that arise from this structure will exhibit multi-collinearity, and when $\psi$ is small compared to the entries in $B$, this multi-collinearity will be very pronounced.  In a factor model, moreover, it need not be the case that $y$ will be associated most strongly with the high-variance principal components of $X$.

We generated data where $p=20$, $n=100$, $k=5$, and $\psi=0.1$, with all the entries of $B$ set to 1.  The resulting coefficient vector, least-squares estimate, and eigenvalues $D$ are excerpted in Table \ref{tab:ridgetrace1}. Principal component 12 is clearly the outlier: it is a strong predictor of $y$, and yet its corresponding variance is two orders of magnitude smaller than the largest variance.

\begin{figure}
\begin{center}
\includegraphics[width=5.0in]{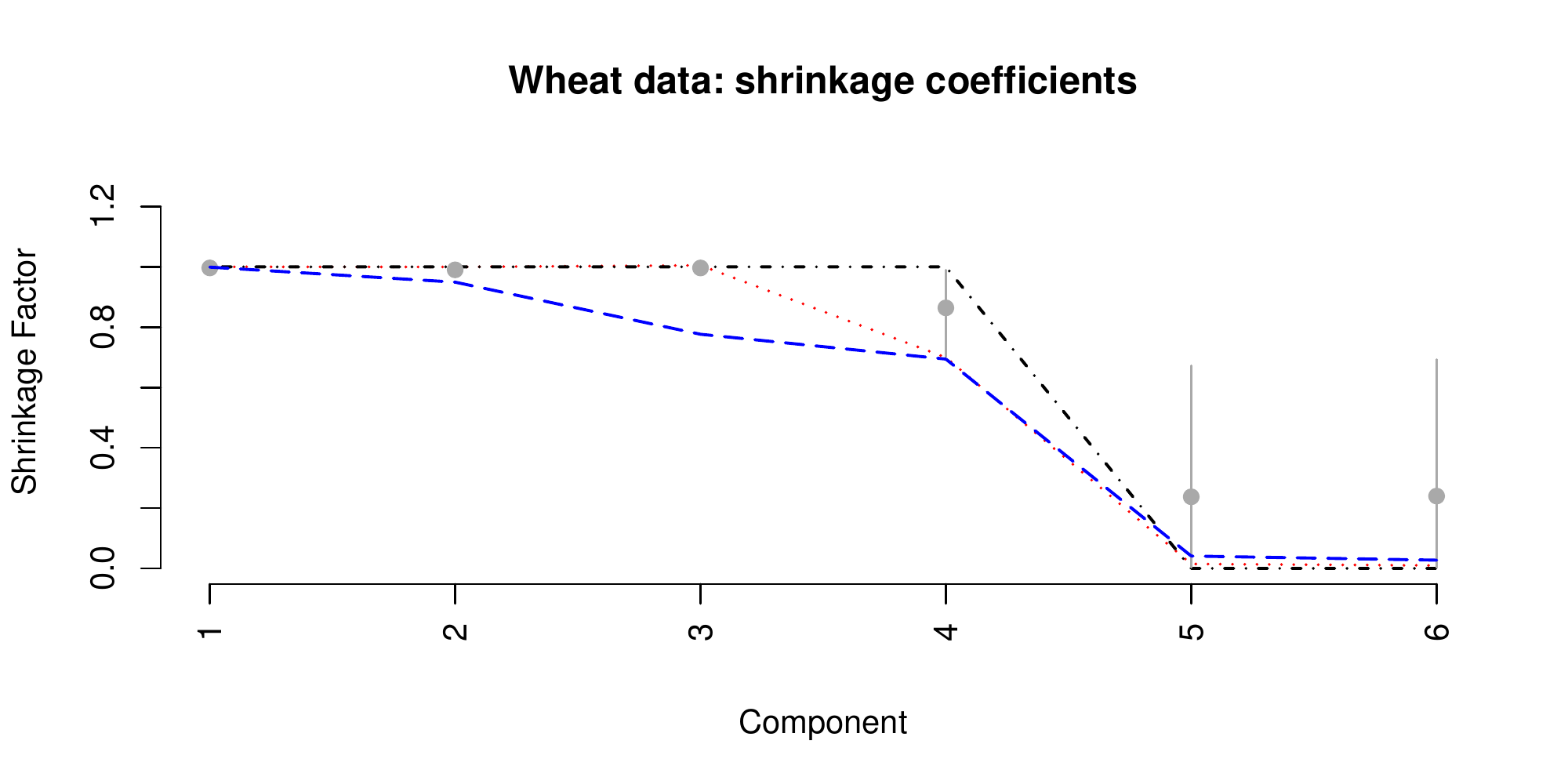} \\
\includegraphics[width=5.0in]{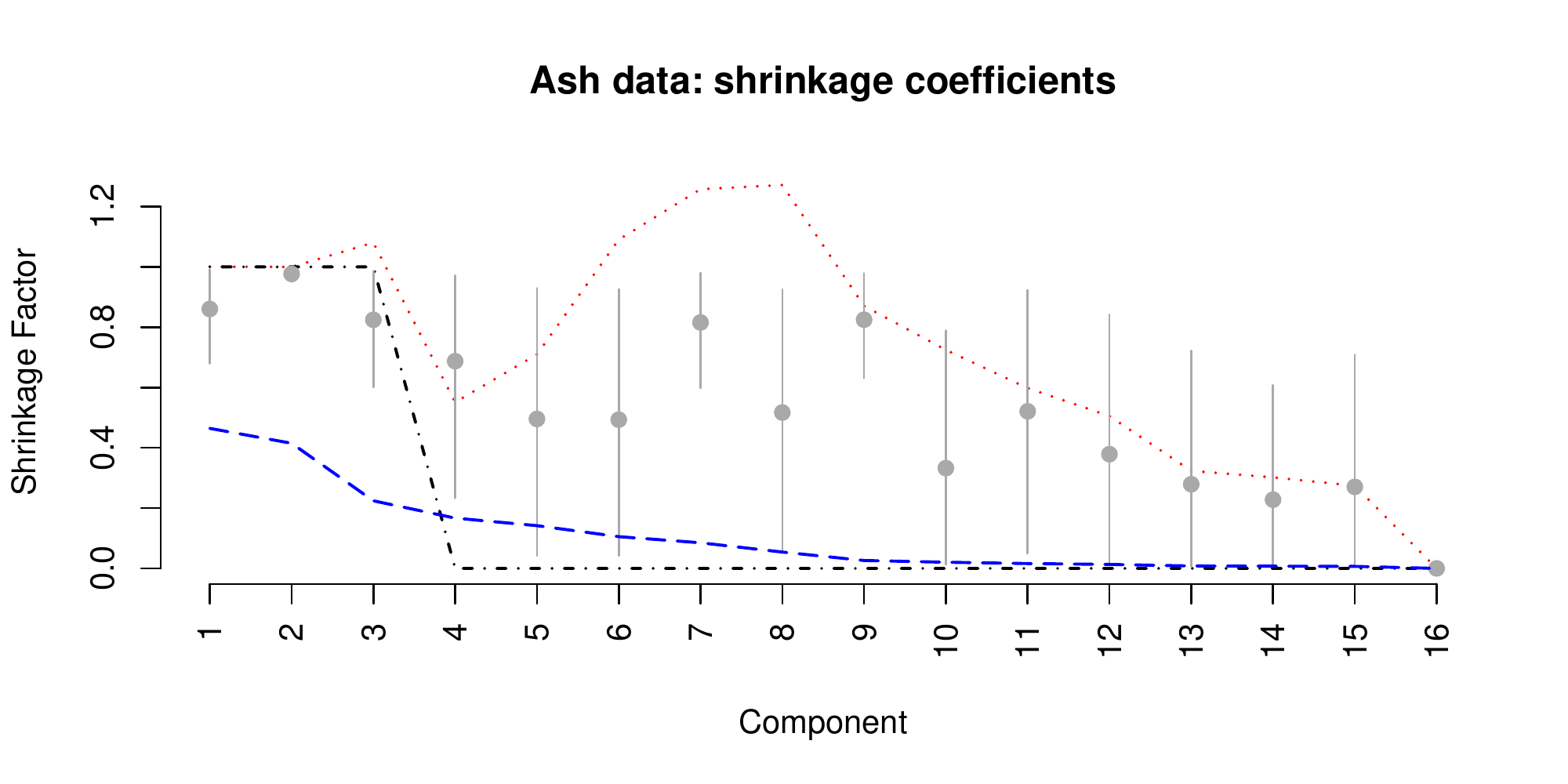} \\
\includegraphics[width=5.0in]{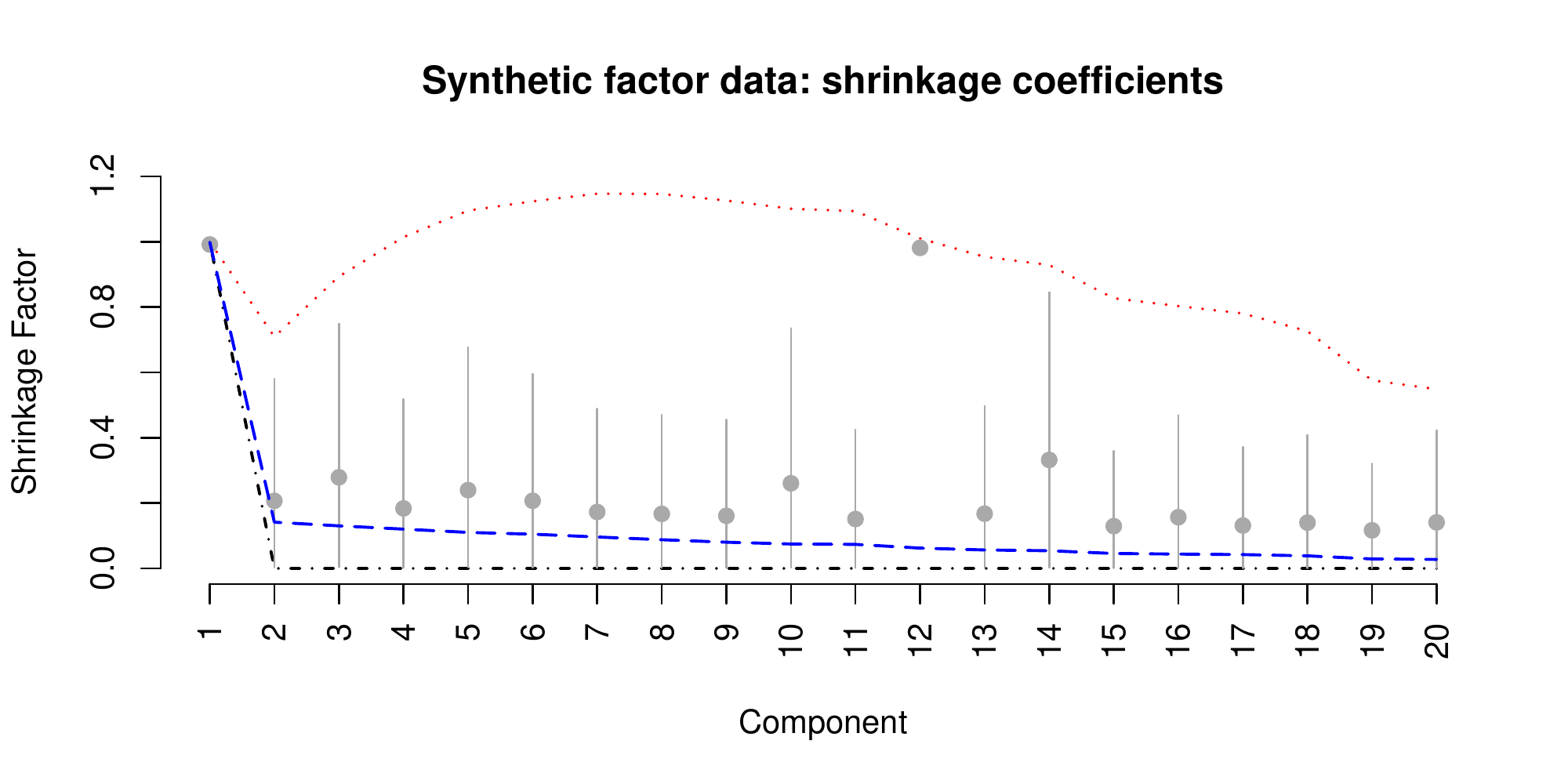}
\caption{\label{fig:examples.kappa}  Comparison on three data sets in terms of how much the four methods shrink each principal component.  Grey dots (grey lines): posterior means (75$\%$ credible intervals) under the fully Bayesian model.  Blue dashes: ridge regression.  Red dots: partial least squares.  Black dots and dashes: principal-component regression.}
\end{center}
\end{figure}

\begin{table}
\begin{center}
\begin{footnotesize}
\caption{\label{tab:ridgetrace1} Subset of the true orthgonalized coefficient vector, least-squares estimate $\hat{\alpha}$, and eigenvalues for Example 3, where $X$ is a five-factor model.}
\vspace{1pc}
\begin{tabular}{r r r r}
Comp. & $\alpha$ & $\hat{\alpha}$ & $D$ \\
\hline
1 & -0.10 & -0.11 & 91.83\\
2 & -0.02 & -0.50 & 1.41\\
$\vdots$ & & & \\
11 & 0.42 & 1.36 & 0.98\\
12 & 12.10 & 12.16 & 0.91\\
13 & 0.04 & 0.13 & 0.85\\
$\vdots$ & & & \\
19 & 0.39 & -1.35 & 0.60\\
20 & 0.00 & -1.87 & 0.58\\
\end{tabular}
\end{footnotesize}
\end{center}
\end{table}

Figure \ref{fig:examples.kappa} compares the shrinkage structures of RR, PCR, PLS, and the Bayesian model for all three of these data sets. The components are ordered left to right along the $x$ axis from highest variance (1) to lowest variance ($p$), while the shrinkage coefficients $\kappa$ (Equation \ref{eqn:shrinkage.general}) are along the $y$ axis.  The tuning parameters for the non-Bayesian methods were chosen by cross-validation.

In all three cases, there appears to be a tendency for both PCR and ridge regression to over-shrink coefficients corresponding to low-variance eigen-directions.  On the ash data set, components 7 and 9 seem to be important, while for the factor model, component 12 is known to be the most important.  Yet all are shrunk nearly to zero by RR and PCR.  For the sake of variance reduction, too much bias is introduced.

Partial least-squares, on the other hand, can identify important low-variance components.  Yet it does so by including many other unimportant low-variance components.  For the sake of bias reduction, too much variance is introduced.

The fully Bayesian model seems to blend the best of both these techniques.  It can successfully pick out important coefficients corresponding to low-variance eigen-directions.  Yet at the same time, it can squelch the other unimportant components.  Intuitively, this combination should make for a favoriable bias--variance tradeoff in larger problems.

\section{Regression when $p>n$}
\label{sec:p>n}

\subsection{Generalization to large-$p$ cases}

Suppose now that the design matrix $X$ is of rank $r<p$ and has singular-value decomposition $X = U D W'$ with $D = \mbox{diag}(d_1, \ldots, d_r)$, again ordered from largest ($d_1$) to smallest ($d_r$).   The approach of the previous section works just as before, with no essential modification:
\begin{eqnarray*}
(\hat{\balpha} \mid \alpha, \sigma^2) &\sim& \N(\alpha, \sigma^2 D^{-2}) \\
(\balpha \mid \sigma^2, \tau^2, \Lambda) &\sim& \N(0, \sigma^2 \tau^2 \Lambda) \\
\lambda_j^2 &\sim& \p(\lambda_j^2) \\
(\sigma^2, \tau^2) &\sim& p(\sigma^2, \tau^2) \, ,
\end{eqnarray*}
where $\balpha = W' \bbeta$ and $\hat{\balpha}$ is the corresponding OLS estimate.  Instead of a $p$-dimensional vector to estimate, we now have an $r$-dimensional one.  Moreover, because we have orthogonalized the coefficients, the elements of $\balpha$ are conditionally independent in the posterior distribution, given $\tau^2$ and $\sigma^2$.  We are faced with a simple normal-means problem, with the only complication being that the singular values $d_j$ enter the likelihood.

This approach is also related to the work of \citet{maruyama:george:2008}, who propose a modification of the standard $g$-prior \citep{zellner86} for use in Bayesian variable selection when $p>n$.  Suppose that
$$
p(\bbeta) = \prod_{j=1}^r p_j(w_j' \bbeta \mid g, \sigma^2) \, .
$$
Each $p_j(w_j' \bbeta \mid g, \sigma^2)$ is a normal density,
\begin{equation}
\label{eqn:MY08prior}
\N \left(w_j' \bbeta \mid 0, \frac{\sigma^2}{d_j^2} f_j (1+g) -\frac{\sigma^2}{d_j^2} \right) \, ,
\end{equation}
where $w_j$ is the $j$th right-singular vector of $X$, and where $f_j > 1$ is necessary to ensure positive definiteness.

The seemingly strange form of (\ref{eqn:MY08prior}) harks back to \citet{strawderman:1971}.  Structurally, it essentially the same prior considered above, with a slight modification made for the sake of ensuring that the marginal distribution $p(\by)$ is analytically convenient \citep[see Section 4.7.10 of][]{bergerbook2ed}.  Maruyama and George recommend mixing over a prior for $g$ while fixing $f_j = d_j^2 / d_r^2$ in (\ref{eqn:MY08prior}).  This approximately corresponds to a similar fixed choice for the $\lambda_j^2$'s in (\ref{eqn:globallocalprior}).

Under this prior, there exists a closed-form expression for the Bayes factor between any two submodels of the full $p$-variable model.  This allows one to perform full Bayesian model selection even when $p>n$. 

Our proposal is an alternative generalization appropriate for pure shrinkage solutions, one that incorporates additional mixing over local variances $\lambda_j^2$. If we treat $W$ as the canonical pseudo-inverse that maps back to the original coordinate system, then the implied prior for $\bbeta = W \alpha$ is a singular normal distribution:
$$
(\bbeta \mid \Lambda, \tau^2, \sigma^2) \sim \N(0, \sigma^2 \tau^2 W \Lambda W') \, .
$$
To see the connection with the $g$-prior more explicitly, suppose that $\lambda_j^2 = d_j^{-2}$ and that $n>p$, such that $X$ is of full column rank.  It is easily verified that $W D^{-2} W' = (X'X)^{-1}$, leading to the original $g$-prior with $g \equiv \tau^2$.  Other authors have considered the same generalization, but with simple conjugate priors for $\lambda_j^2$---for example, \citet{clydeJASA1996}, \citet{denison:george:2000}, and \citet{west03}.  Our approach differs in our emphasis placed upon the choice of prior for $\lambda_j^2$, for which the developments earlier in the paper are clearly relevant.

Under this model, the (conditional) posterior mean estimator for $\alpha_j$ is, just as before, given by
$$
\left( \frac{\tau^2 \lambda_j^2 d_j^2}{1 + \tau^2 \lambda_j^2 d_j^2} \right) \hat{\alpha}_j \, ,
$$
a generalized Bayesian version of the classic ridge estimator.

\subsection{Assessing out-of-sample predictive performance}

In the following simulation studies, we investigate the performance of the Bayesian model proposed above.  We use the horseshoe prior, whereby $\tau$ and each $\lambda_j$ receive independent half-Cauchy priors.  We now sketch a brief rationale for this choice.  Intuitively, the vectors $\{w_j\}$ can be thought of as contrasts.  A nice ``default'' Bayesian model would express the prior belief that certain contrasts of the $\bbeta$ sequence will be strong predictors of $\by$, and that some will be weak predictors.  The horseshoe prior does just this: it will shrink most $\alpha_j$'s very strongly, as the posterior mass for $\tau$ tends to concentrate near zero.  Yet it will leave unshrunk those $\alpha_j$'s corresponding to contrasts that predict $\by$ well---even, it is to be hoped, those that correspond to a low-variance principal components---since the heavy tails of the half-Cauchy prior will allow certain $\lambda_j$'s to be quite large.

As test cases, we used the following 7 data sets, all of which had more predictors than observations.  Only 1 of the 7 data sets is simulated; the other 6 are from chemometrics or genomics.  All are available upon request from the authors, and the 6 real data sets are available from the R packages \verb|pls|, \verb|chemometrics|, and \verb|mixOmics|.
\begin{description}
\item[factor:] the only simulated data set considered.  Both $X$ and $y$ were generated jointly from a standard Bayesian factor model, with $y$ loading most heavily on the lowest-variance factors.
\item[nutrimouse:] observations of 40 mice where hepatic fatty-acid concentrations are regressed upon the expression of 120 potentially relevant genes measured in liver cells.
\item[cereal:] chemometric observations of 15 cereal molecules where starch content is regressed upon NIR spectra at 145 different wavelengths.
\item[yarn:] samples of 28 polyethylene terephthalate (PET) yarns, where the density of the yarn sample is regressed upon measurements of NIR spectra at 268 wavelenths.
\item[gasoline:] octane numbers of 60 gasoline samples along with NIR spectra at 401 wavelengths.
\item[multidrug:] the $X$ matrix comprises observations of the activity of 853 drugs on 60 different human cell lines, expressed as the concentration at which each drug leads to a $50\%$ inhibition of growth for each cell line.  The $y$ variable is the measured expression of ABC3A (an ATP-binding cassette transporter) in each cell line.
\item[liver:] the $X$ matrix contains the expression scores for 3116 genes in 64 rat subjects.  The $y$ variable is the cholesterol concetration in the liver.
\end{description}

\begin{table}
\begin{center}
\caption{\label{tab:holdoutresults} Average out-of-sample predictive error (SSE) on 50 different train/test splits for 7 data sets where $p > n$.  Bayes: the local-shrinkage model with horseshoe priors.  PLS: partial least squares.  PCR: principal-components regression.  RR: ridge regression.  SPLS: sparse partial least squares.  The smallest entry in each row is in boldface.}
\vspace{1pc}
\begin{tabular}{r r r r r r r r r}
& & & & \multicolumn{5}{c}{Average out-of-sample error} \\
Data set & $n$ & $p$ & & Bayes & PLS & PCR & RR & SPLS \\
\hline
factor & 50 & 100 & & \textbf{45.8} & 66.9 & 69.2 & 358 &  97.6 \\
nutrimouse & 40 & 120 & & \textbf{394} & 428 & 467 & \textbf{394} &  462 \\
cereal & 15 & 145 & & 45.2 & 46.9 & 46.3 & \textbf{42.2} &  46.5 \\
yarn & 28 & 268 & & \textbf{2.63} & 6.89 & 20.2 & 4.18 &  53.8 \\
gasoline & 60 & 401 & & 0.82 & 0.87 & 0.93 & \textbf{0.72} &  1.04 \\
multidrug & 60 & 853 & & \textbf{139} & 152 & 173 & 143 &  160 \\
liver & 64 & 3116 & & \textbf{1340} & 1457 & 1475 & 1407 &  1470
\end{tabular}
\end{center}
\end{table}

We compare the Bayesian model to the three basic techniques (partial least squares, ridge regression, and principal-components regression), along with a new technique called sparse partial least squares \citep{chun:keles:2010} aimed at simultaneous dimension reduction and variable selection.  This final method is implemented in the R package \verb|spls|.

To test these five methods, we split each of the seven data sets into training and test samples, with $75\%$ of the observations used for training.  We then fit each model using the training data, with tuning parameters for the non-Bayesian methods chosen by ten-fold cross validation on the training data alone.  We then compared out-of-sample predictive performance on the holdout data, measured by sum of squared prediction errors (SSE).  In each case the $y$ variable was centered, and the $X$ variables were centered and scaled.

All of our results in Table \ref{tab:holdoutresults} represent the average SSE incurred over 50 different train/test splits.  There are several interesting things to notice here.  For one thing, the Bayes method seems to be the overall winner.  It was the outright best on 4 data sets, tied for best on 1 data set, and second-best on the other two data sets.  Surprisingly, the next-best method seems to be a venerable classic: ridge regression.  The newest method, sparse partial least squares, was either worst or second-worst on all 7 data sets.

The two cases where the Bayesian method offered the biggest improvements---the factor data and the yarn data---are also instructive.  In these cases, the $y$ variable was most strongly associated with smaller-variance contrasts $w_j$, or in other words, those contrasts associated with smaller singular values $d_j$.  Much as we saw in the previous section, classic methods like ridge regression and PCR perform poorly when this is the case, whereas the Bayesian model is quite robust.

In other cases (notably the cereal, gasoline, and nutrimouse data sets), the signal-to-noise ratio seems to be either so favorable, or so poor, that all the methods do almost equally well.  This suggests that the extra variance induced by mixing over local $\lambda_j^2$'s does not pose difficulty for the Bayesian model.

\section{Final Remarks}

The study of oracle properties provides a unifying framework in the classical literature for the study of regularized regression, but no such framework exists for Bayesians.  In this paper, we have offered a few elements that might form the beginnings of such a framework.  By identifying $\bbeta$ (or $\balpha$) with the increments of a discretely observed L\'evy process, we have embedded the finite-dimensional problem in a suitable infinite-dimensional generalization.  This provides a natural setting in which the dimension $p$ grows without bound.  In particular, Theorem \ref{the:subordinatorpenalty} establishes mappings among L\'evy processes, penalty functions, priors, and scale mixtures of well-known distributions.  This offers a convenient way of generating infinitely divisible probability distributions with known probabilistic structure, giving Bayesian statisticians a much larger toolbox for building shrinkage models like the kind explored in Section \ref{sec:p>n}.

\appendix

\section{Proofs of main results}
\label{app:proofs}

\subsection*{Proof of Theorem \ref{the:subordinatorpenalty}}

For Part A, since $\psi(t)$ is totally monotone, it has derivatives of all orders and satisfies
$$
(-1)^{n} \psi^{(n)}(t) \leq 0 \, .
$$
Furthermore, since $\lim_{t \to 0} \psi(t) = 0$, then by Bernstein's theorem $\psi(t)$ corresponds to the Laplace exponent of some subordinator $T(s)$ \citep[see, e.g.,][Chapter 4]{cont:tankov:2004}.  That is, there exists a subordinator $T(s)$ with L\'evy measure $\mu(dx)$ whose moment-generating function can be written as
\begin{equation}
\label{eqn:subordinatormgf}
M_{T}(t) = \E \{ \exp[-t T(s)] \} = \exp \{ -s \psi(t) \} \, ,
\end{equation}
where $\psi(t)$ is called the Laplace exponent and is given by its L\'evy representation in Equation (\ref{eqn:levyrepsubordinator}).

We recognize the mixture-of-normals representation in Part B as follows.  Write the expectation in (\ref{eqn:subordinatormgf}), evaluated at $t = \beta_j^2/2$, as
\begin{eqnarray*}
\E \{ \exp(-t T_s \} &=& \int_0^{\infty} \exp \{ -\beta_j^2 T_s/2 \} \ p(T_s) \dd T_s \\
&=& \int_0^{\infty} \sqrt{T_s} \exp \big\{ -\beta_j^2 T_s /2\} \ \{ T_s^{-1/2} p(T_s) \big\} \ \dd T_s \, ,
\end{eqnarray*}
where $p(T_s)$ is the marginal density of the subordinator $T$ observed at time $s$.  The expression $T_s^{-1/2} p(T_s)$ is thus clearly proportional to a prior density for the precision $T$ in a Gaussian mixture for $\beta_j^2$.

This gives an explicit representation of the mixing density as the power-tilted density of the subordinator when $\alpha = 2$.


\subsection*{Proof of Theorem \ref{thm:masreliezextension}}

By definition, $p( \beta ) = \int_0^\infty e^{ - T \frac{\beta^2}{2} } g(T)  d T  $. Therefore 
$$
m( y) = \int p( y - \beta ) \int_0^\infty e^{ - T \frac{\beta^2}{2} } g ( T ) d T d \beta \, .
$$
The posterior mean is given by
\begin{align*}
E( \beta | y ) & = \frac{1}{m(y)} \int p( y - \beta ) \beta e^{ - T \frac{\beta^2}{2} } g ( T ) d T d \beta \\
& = \frac{1}{m(y)} \int p( y - \beta ) d \left ( - e^{ - T \frac{\beta^2}{2} } \right ) T^{-1} g ( T ) d T d \beta \, .
\end{align*}
Using integration by parts yields
\begin{align*}
E( \beta | y ) & =  \frac{ E(T^{-1} )}{m(y)}
 \int  \frac{\partial}{\partial y} p( y - \beta ) e^{ - T \frac{\beta^2}{2} } g^\star ( T ) d T d \beta \\
 & = E(T^{-1}) \frac{m^\star (y)}{m(y)} \frac{\partial}{\partial y} \ln m^\star ( y) \, .
\end{align*}

\begin{small}
\singlespace
\bibliographystyle{abbrvnat}
\bibliography{masterbib}
\end{small}


\end{document}